\documentclass[11pt]{article}
\usepackage{fancyhdr}

\usepackage{url,hyperref,lineno,microtype,subcaption}
\usepackage{amssymb}
\usepackage{setspace}
\usepackage{graphics,epsfig,cite,amsmath,amsthm}
\usepackage[all]{xy}

\usepackage{algorithm}
\usepackage[noend]{algpseudocode}
\usepackage{multirow}

\providecommand{\keywords}[1]{\textbf{\textit{Keywords:---}} #1}

\newcommand{\pu}{p_k^{\uparrow}}
\newcommand{\pd}{p_k^{\downarrow}}

\newcommand{\Expect}{{\rm I\kern-.3em E}}

\newtheorem{theorem}{Theorem}[section]   

\newtheorem{example}[theorem]{Example}        

\begin{document}

\title{Estimating Propensity Parameters using Google PageRank and Genetic Algorithms}

\author{David Murrugarra$^{1}$\thanks{Correspondence: \href{murrugarra@uky.edu}{murrugarra@uky.edu}} \and Jacob Miller$^{1}$ \and Alex Mueller$^{1}$}
\date{}
\maketitle
\thispagestyle{fancy} 
{\footnotesize
  \centerline{$^{1}$Department of Mathematics,
  University of Kentucky, Lexington, KY 40506-0027 USA.}
}
\begin{abstract} 
Stochastic Boolean networks, or more generally, stochastic discrete networks, are an important class of computational models for molecular interaction networks. The stochasticity stems from the updating schedule. Standard updating schedules include the synchronous update, where all the nodes are updated at the same time, and the asynchronous update where a random node is updated at each time step. The former produces a deterministic dynamics while the latter a stochastic dynamics. A more general stochastic setting considers propensity parameters for updating each node. Stochastic Discrete Dynamical Systems (SDDS) are a modeling framework that considers two propensity parameters for updating each node and uses one when the update has a positive impact on the variable, that is, when the update causes the variable to increase its value, and uses the other when the update has a negative impact, that is, when the update causes it to decrease its value. This framework offers additional features for simulations but also adds a complexity in parameter estimation of the propensities. This paper presents a method for estimating the propensity parameters for SDDS. The method is based on adding noise to the system using the Google PageRank approach to make the system ergodic and thus guaranteeing the existence of a stationary distribution. Then with the use of a genetic algorithm, the propensity parameters are estimated. Approximation techniques that make the search algorithms efficient are also presented and Matlab/Octave code to test the algorithms are available at~\href{http://www.ms.uky.edu/~dmu228/GeneticAlg/Code.html}{http://www.ms.uky.edu/$\sim$dmu228/GeneticAlg/Code.html}.
\end{abstract}
\section{Introduction}
Mathematical modeling has been widely applied to the study of biological systems with the goal of understanding the important properties of the system and to derive useful predictions about the system. The type of systems of interest ranges from the molecular to ecological systems. At the cellular level, gene regulatory networks (GRN) have been extensively studied to understand the key mechanisms that are relevant for cell function. GRNs represent the intricate relationships among genes, proteins, and other substances that are responsible for the expression levels of mRNA and proteins. The amount of these gene products and their temporal patterns characterize specific cell states or phenotypes~\cite{Murrugarra:2015aa}.

Gene expression is inherently stochastic with randomness in transcription and translation. This stochasticity is usually referred to as noise and it is one of the main drivers of variability~\cite{ Raj:qf}. Variability has an important role in cellular functions, and it can be beneficial as well as harmful~\cite{Eldar:2010kq,Kaern:2005qv}. Modeling stochasticity is an important problem in systems biology. Different modeling approaches can be found in the literature. Mathematical models can be broadly divided into two classes: continuous, such as systems of differential equations and discrete, such as Boolean networks and their generalizations. This paper will focus on discrete stochastic methods. The Gillespie algorithm~\cite{ Gillespie77, doi:10.1146/annurev.physchem.58.032806.104637} considers discrete states but continuous time. In this work, we will focus on models where the space as well as the time are discrete variables. For instance, Boolean networks (BNs) are a class of computational models in which genes can only be in one of two states: ON or OFF. BNs and, in general, multistate models, which allow genes to take on more than two states, have been effectively used to model biological systems such as the \textit{p53-mdm2} system~\cite{Choi2012,Murrugarra2012,Abou-Jaoude:2009aa}, the \textit{lac} operon~\cite{DBLP:journals/jcb/Veliz-CubaS11}, the yeast cell cycle network~\cite{Li:2004aa}, the Th regulatory network~\cite{Mendoza:2006aa}, A. \textit{thaliana}~\cite{Balleza:2008aa}, and many other systems~\cite{Davidich:2008aa, Albert:2003aa, Saadatpour:2011aa, Zhang:2008aa, Helikar:2008aa, Helikar:2013aa}.

Stochasticity in Boolean networks has been studied in different ways. The earliest approach to introduce stochasticity into BNs was the asynchronous update, where a random node is updated at each time step~\cite{Thomas:1990aa}. Another approach considers update sequences that can change from step to step~\cite{ Saadatpour:2010aa, Mortveit:2007:ISD:1201984}. More sophisticated approaches include Probabilistic Boolean Networks (PBNs)~\cite{Shmulevich2002} and their variants~\cite{Layek:2009pi,Liang:2012la}. PBNs consider stochasticity at the function level where each node can use multiple functions with a switching probability from step to step. SDDS~\cite{Murrugarra2012} is a simulation framework similar to PBNs but the key difference is how the transition probabilities are calculated. SDDS considers two propensity parameters for updating each node. These parameters resemble the propensity probabilities in the Gillespie algorithm~\cite{Gillespie77, doi:10.1146/annurev.physchem.58.032806.104637}. This extension gives a more flexible simulation setup as a generative model but adds the complexity of parameter estimation of the propensity parameters. This paper provides a method for computing the propensity parameters for SDDS. 

For completeness, in the following subsection, we will define the stochastic framework to be used in remainder of the paper.
\subsection*{Stochastic Framework}
In this paper we will focus on the stochastic framework introduced in~\cite{Murrugarra2012} referred to as Stochastic Discrete Dynamical Systems (SDDS). This framework is a natural extension of Boolean networks and is an appropriate setup to model the effect of intrinsic noise on network dynamics. Consider the discrete variables $x_1, \ldots , x_n$ that can take values in finite sets $S_1,\ldots , S_n$, respectively. Let $S = S_1\times\cdots\times S_n$ be the Cartesian product. A \emph{SDDS} in the variables $x_1, \ldots , x_n$ is a collection of $n$ triplets
\begin{displaymath}
F=\{f_i,p_i^\uparrow,p_i^\downarrow\}^n_{i=1}
\end{displaymath}
where  
\begin{itemize}
  \item $f_i : S\rightarrow S_i$ is the update function for $x_i$, for all $i = 1,\dots,n$.
  \item $p_i^\uparrow$ is the activation propensity.
  \item $p_i^\downarrow$ is the degradation propensity.
  \item $p_i^\uparrow,p_i^\downarrow\in[0,1]$. These are the parameters of interest in this paper.
\end{itemize}

The stochasticity originates from the propensity parameters $\pu$ and $\pd$, 
which should be interpreted as follows: If there would be an activation of $x_k$
at the next time step, 
i.e., if $s_1,s_2\in S_k$ with $s_1<s_2$ and $x_k(t)=s_1$, and $f_k(x_1(t),\ldots, x_n(t)) = s_2$, then $x_k(t+1) = s_2$ with probability $\pu$. The degradation probability $\pd$ is defined similarly.
SDDS can be represented as a Markov chain by specifying its transition matrix in the following way. For each variable $x_i$, $i=1,\dots,n$, the probability of changing its value is given by
\[
Prob(x_i\rightarrow f_i(x))=
\begin{cases}
p_i^\uparrow, & \text{if $x_i<f_i(x)$},\\
p_i^\downarrow, & \text{if $x_i>f_i(x)$},\\
1, & \text{if $x_i=f_i(x)$},
\end{cases}
\]
and the probability of maintaining its current value is given by
\[
Prob(x_i\rightarrow x_i)=
\begin{cases}
1-p_i^\uparrow, & \text{if $x_i<f_i(x)$},\\
1-p_i^\downarrow, & \text{if $x_i>f_i(x)$},\\
1, & \text{if $x_i=f_i(x)$}.
\end{cases}
\]

Let $x,y\in S$. The transition from $x$ to $y$ is given by
\begin{equation}\label{transition_x_y}
a_{xy}=\prod_{i=1}^n Prob(x_i\rightarrow y_i).
\end{equation}
Notice that $Prob(x_i\rightarrow y_i)=0$ for all $y_i\notin \{x_i,f_i(x)\}$.

Then the transition matrix is given by
\begin{equation}\label{transition_matrix_SDDS}
A = (a_{xy})_{x,y\in S}
\end{equation}
The dynamics of SDDS depends on the transition probabilities $a_{xy}$, which depend on
the propensity values and the update functions. Online software to test examples is available at \url{http://adam.plantsimlab.org/} (choose Discrete Dynamical Systems (SDDS) in the model type).

In Markov chain notation, the transition probability $a_{xy} = p(X_t=x|X_{t-1}=y)$ represents the probability of being in state $x$ at time t given that system was in state $y$ at time $t-1$. If $\pi_t=p(X_t=x)$ represents the probability of being in state $x$ at time $t$, then we will assume that \textbf{$\pi$} is a row vector containing the probabilities of being in state $x$ at time $t$ for all $x\in S$. If \textbf{$\pi_0$} is the initial distribution at time $t=0$, then at time $t=1$,
\begin{equation}
\label{Eq:trans_pi}
\pi_1=\sum_{x\in S}\pi_0(x)a_{xy}.
\end{equation}

If we iterate Equation~\ref{Eq:trans_pi} and if we get to the point where  
\begin{equation}
\label{Eq:stationary_dist}
\pi=\sum_{x\in S}\pi(x)a_{xy}
\end{equation}
then we will say that the Markov chain has reached a stationary distribution and that $\pi$ is the stationary distribution.
\section{Methods}
In this section we describe a method for estimating the propensity parameters for SDDS.
The approach is based on adding noise to the system using the Google PageRank~\cite{Brin20123825, DavidLay, murphy2012machine} strategy
to make the system ergodic and thus guaranteeing the existence of a stationary distribution and then with the use of a genetic algorithm the propensity parameters are estimated. To guarantee the existence of a stationary distribution we use a special case of the Perron-Frobenius Theorem. 
\begin{theorem}[Perron-Frobenius]\label{PerronFrobenius}
If $\textbf{A}$ is a regular $m\times m$ transition matrix with $m\geq2$, then
\begin{itemize}
\item For any initial probability vector $\pi_0$, $\lim_{n\rightarrow\infty}\textbf{A}^n\pi_0 = \pi$.
\item The vector $\pi$ is the unique probability vector which is an eigenvector of $\textbf{A}$ associated with the eigenvalue 1.
\end{itemize}
\end{theorem}
A proof of Theorem~\ref{PerronFrobenius} can be found in Chapter 10 of~\cite{DavidLay}. 

Theorem~\ref{PerronFrobenius} ensures a unique stationary distribution $\pi$ provided that we have a regular transition matrix, that is, if some power $\textbf{A}^k$ contains only strictly positive entries. However, the transition matrix $\textbf{A}$ of SDDS given in Equation~\ref{transition_matrix_SDDS} might not be regular. 
In the following subsection, we use a similar approach to the Google's PageRank algorithm to add noise to the system to obtain a new transition matrix that is regular.
\subsection{PageRank Algorithm}
For simplicity, consider a SDDS, $F=\{f_i,p_i^\uparrow,p_i^\downarrow\}^n_{i=1}$ where $f_i : S\rightarrow S_i$, $S=\mathbb{K}^n$, and $|\mathbb{K}|=p$. Then its transition matrix $\textbf{A}$ given in Equation~\ref{transition_matrix_SDDS} is a $p^n\times p^n$ matrix. To introduce noise into the system we
consider the Google Matrix
\begin{equation}
\label{Eq:GoogleMatrix}
\textbf{G} = g\textbf{A}+(1-g)\textbf{K},
\end{equation}
where $g$ is a constant number in the interval $[0,1]$ and $\textbf{K}$ is a $p^n\times p^n$ matrix all of whose columns are the vector $(1/p^n,\dots,1/p^n)$.
The matrix $\textbf{G}$ in Equation~\ref{Eq:GoogleMatrix} is a regular matrix and then we can use Theorem~\ref{PerronFrobenius} to get a stationary distribution for $\textbf{G}$,
\begin{equation}
\label{Eq:StationaryDist}
\pi = \pi_{\textbf{G}} = (\pi_1,\dots,\pi_{p^n})
\end{equation}
This stationary distribution reflects the dynamics of the SDDS $F=\{f_i,p_i^\uparrow,p_i^\downarrow\}^n_{i=1}$. The importance of a state $x\in S$ can be measured by the size of the corresponding entry $\pi_x$ in the stationary distribution of Equation~\ref{Eq:StationaryDist}. For instance, for ranking the importance of the states in a Markov chain one can use the size of the corresponding entries in the stationary distribution. We will refer to this entry $\pi_x$ as the PageRank score of $x$.
\subsection{Genetic Algorithm}
The entries of the stationary distribution $\pi$ in Equation~\ref{Eq:StationaryDist} can also be interpreted as occupation times for each state. Thus it gives the probability of being at a certain state. Now suppose that we start with a desired stationary distribution $\pi^\ast = (\pi^\ast_1,\dots,\pi^\ast_{p^n})$. We have developed a genetic algorithm that initializes a population of random propensity matrices and searches for a propensity matrix $prop^\ast$ such that its stationary distribution $\pi = (\pi_1,\dots,\pi_{p^n})$ gets closer to the desired stationary distribution $\pi^\ast$. That is, we search for propensity matrices such that the distance between $\pi$ and $\pi^\ast$ is minimized,
\begin{equation}
\label{Eq:min_dist}
\min_{p_i^\uparrow,p_i^\downarrow}d(\pi,\pi^\ast)\ \text{ or }\ \min_{p_i^\uparrow,p_i^\downarrow}|\pi(j)-\pi^\ast(j)|
\end{equation}
The pseudocode of this genetic algorithm is given in Algorithm~\ref{GeneticAlg} and it has been implemented in Octave/Matlab and our code can be downloaded from~\href{http://www.ms.uky.edu/~dmu228/GeneticAlg/Code.html}{http://www.ms.uky.edu/$\sim$dmu228/GeneticAlg/Code.html}.
\begin{algorithm}[H]
\caption{Genetic Algorithm with PageRank.}\label{GeneticAlg}
\algrenewcommand\algorithmicforall{\textbf{for each}}
\begin{algorithmic}[1]
\Require Functions: $F=(f_1,\dots,f_n)$, number of generations: $NumGen$, population size: $PopSize$, states of interest: $States$, and desired probabilities: $\pi^\ast=\pi^\ast(States)$.
\Ensure Propensity parameters: $\textbf{prop}^\ast$ 
\Procedure {GeneticGoogle}{$F$, $NumGen$, $PopSize$, $\pi^\ast$}
\State $PopPropensities\leftarrow$ initialize a population of propensity matrices.
\State [$fitnesses$,$\min(PopPropensities)$] = \Call{FitnessGoogle}{$F$, $PopPropensities$, $\pi^\ast$}
\For{i=1,\dots, $NumGen$}
	\State $NewPropensities\leftarrow$ initialize new population of propensities.
    	\For{j=1,\dots, $PopSize$}
		 \If {$rand < fitnesses(j)$ }
			\State $parent1(j) = PopPropensities(j)$
		\Else	
			\State $parent2(j) = PopPropensities(j)$
  		\EndIf
    		\State $children$ = \Call{Crossover}{$parent1$, $parent2$, $mut$, $\sigma$} 
    		\State $NewPropensities(j) = children$ 
    \EndFor
    \State [$fitnesses$,$\min(NewPropensities)$] = \Call{FitnessGoogle}{$F$, $NewPropensities$, $\pi^\ast$} 
    \State $PopPropensities = NewPropensities$.
 \EndFor
\State $\textbf{prop}^\ast = \min(NewPropensities)$.
\EndProcedure
\Function{FitnessGoogle}{$F$, $PopPropensities$, $\pi^\ast$}
\For{i = 1,\dots, length($PopPropensities$)}\Comment{For each propensity matrix.} 
\State $\pi =
\begin{cases}
      PageRank(F,PopPropensities(i))& \text{for exact distribution, see Equation~\ref{Eq:StationaryDist}}, \\
      \Call{EstimateStaDist}{ F, c, NumIter, g}& \text{for estimated distribution, see Algorithm~\ref{EstimateStaDist}}.
\end{cases}$
\State $d = d(\pi,\pi^\ast)$\Comment{We used a weighted distance to give predominance to important states.}
\State $fitnesses(i) = e^{(-d^2/s)}$
\EndFor
\State \Return([$fitnesses$,$\min(PopPropensities)$])\Comment{Keep propensity with minimum fitness.}
\EndFunction
\Function{Crossover}{$parent1$, $parent2$, $mut$, $\sigma$}
\State $NewProp\leftarrow$ initialize new propensity matrix.
\State $DivLine = $ random integer between 1 and length($parent1$).
\For{i = 1,\dots, length($parent1$)}
 	\If {$i < DivLine$ }
		\State $NewProp(i) = parent1(i)$
	\Else	
		\State $NewProp(i) = parent2(i)$
  	\EndIf
	\If{$rand <mut$}
		\State $NewProp(i) = NewProp(i)+ normrand(0,\sigma)$ \Comment{Introduce mutation.}
	\EndIf 
\EndFor
\State \Return ($NewProp$)
\EndFunction 
\end{algorithmic}
\end{algorithm}
\subsection{Estimating the stationary distribution}
The genetic algorithm, Algorithm~\ref{GeneticAlg}, uses the exact stationary distribution through PageRank (see Equation~\ref{Eq:StationaryDist}) which is computationally expensive for larger models. Here we present an efficient algorithm for estimating the stationary distribution based on a random walk.
The expensive part of Algorithm~\ref{GeneticAlg} is the calculation of the stationary distribution $\pi$ in Equation~\ref{Eq:StationaryDist}.
We have implemented an algorithm for estimating the stationary distribution by doing a random walk using SDDS as a generative model; see Algorithm~\ref{EstimateStaDist}. The idea behind Algorithm~\ref{EstimateStaDist} is to use SDDS for simulating from an initial state according to the transition probabilities given in Equation~\ref{Eq:GoogleMatrix}. That is, we initialize the simulation at an initial state $x\in S$ and then with probability $g$
we move to another state $y\in S$ according to $a_{xy}$ (see Equation~\ref{transition_x_y}) and with probability $1-g$ we jump to a random node. We repeat this process for a given number of iterations. At the end of a maximum number of iterations, we count how often we visited each state and the normalized frequencies will be the approximated stationary distribution.

To make the genetic algorithm more efficient, we used the estimated stationary distribution described in the previous paragraph.
Thus, within the fitness function of Algorithm~\ref{GeneticAlg}, we use the estimated stationary distribution to
assess the fitness of the generated propensity matrices. The pseudocode for this new algorithm is the same as Algorithm~\ref{GeneticAlg}, the only change is in the fitness function.
This version of the algorithm has also been implemented in Octave/Matlab and our code can be found in~\href{http://www.ms.uky.edu/~dmu228/GeneticAlg/Code.html}{http://www.ms.uky.edu/$\sim$dmu228/GeneticAlg/Code.html}.
\begin{algorithm}
\begin{algorithmic}[2]
\caption{Estimate Stationary Distribution.}\label{EstimateStaDist}
\Require Functions: $F=(f_1,\dots,f_n)$, propensities: $c$, number of iterations: $NumIter$, noise: $g$.
\Ensure Estimated stationary distribution $\pi$
\State $\pi = \Call{EstimateStaDist}{ F, c, NumIter, g}$
\State \Return $\pi$ 
\Function{EstimateStaDist}{$F, c, NumIter$, $g$}
\State $distribution\leftarrow$ initialize frequency vector.
\State $s\leftarrow$ initialize random initial state.
\For{ i=1,\dots, $NumIter$ }
 	\If {$rand < g$ }
		\State $y = $ random state between 1 and $p^n$.
	\Else	
    		\State $y = SDDS.\text{nextstate}(s, c ) $ 
  	\EndIf
\State $distribution(y) =+ 1$ increase state frequency.
\State $sum = $ total frequencies. 
\State $ \pi = distribution/sum $
\EndFor
\State \Return $\pi$     
\EndFunction
\end{algorithmic}
\end{algorithm}
\section*{Results}
We test our methods using two published models that are appropriate for changing the stationary distribution under the choice of different propensity parameters. The first model is a Boolean network while the second is a multistate model. In both models bistability has been observed but the basin size of one of the attractors under the synchronous update is much larger than the basin of the other attractor, and thus the stationary distribution will be more concentrated in one of the attractors. We will use our methods to change the stationary distribution in favor of the attractor with a smaller basin.
\begin{example}\label{lac_operon} \textbf{Lac-operon network.}
The lac-operon in \textit{E.~coli}~\cite{JACOB1961318} is one of the best studied gene regulatory networks. This system is responsible for the metabolism of lactose in the absence of glucose. This system exhibits bistability in the sense that the operon can be either ON or OFF, depending on the presence of the preferred energy source: glucose. A Boolean network for this system has been developed in~\cite{DBLP:journals/jcb/Veliz-CubaS11}. 
This model considers the following 10 components
\begin{equation}\label{Eq:lacten_node_labels}
\begin{tabular}{ll}
$x_1 = $ \textit{M}: \textit{lac} mRNA, & $x_2 = $ \textit{P}: \textit{lac permease}, \\
$x_3 = $ \textit{B}: \textit{lac}$\beta$-\textit{galactosidase}, &$x_4 = $ \textit{C}: CAP, \\
$x_5 = $ \textit{R}: repressor, &$x_6 = $ \textit{Rm}: repressor at medium concentration,\\
$x_7 = $ \textit{A}: allolactose, &$x_8 = $ \textit{Am}: allolactose at medium concentration, \\
$x_9 = $ \textit{L}: lactose, &$x_{10} = $ \textit{Lm}: lactose at medium concentration, \\
\end{tabular}
\end{equation}
and the Boolean rules are given by
\begin{equation}\label{Eq:T-GLG-polynomials}
\begin{tabular}{l}
$f_1=x_4\wedge\overline{x_{5}}\wedge\overline{x_{6}}$,\\
$f_2=x_1$,\\
$f_3= x_1$,\\
$f_4= \overline{G_e}$,\\
$f_5= \overline{x_{7}}\wedge\overline{x_{8}}$,\\
$f_6= (\overline{x_{7}}\wedge\overline{x_{8}})\vee x_5$,\\
$f_7= x_{9}\wedge x_{3}$,\\
$f_8= x_{9}\vee x_{10}$,\\
$f_9= x_2\wedge L_e\wedge\overline{G_e}$,\\
$f_{10}= ((L_{em}\wedge P)\vee L_e) \wedge\overline{G_e}$.
\end{tabular}
\end{equation}
where $G_e$, $L_{em}$, and $L_e$ are parameters. $G_e$ and $L_e$ indicate the concentration of extracellular glucose and lactose, respectively.
The parameter $L_{em}$ indicates the medium concentration of extracellular lactose.
For medium extracellular lactose, that is when $L_{em}=1$ and $L_e=0$, this system has two fixed points: $s_1 = (0, 0, 0, 1, 1,1, 0, 0, 0, 0)$ and $s_2=(1, 1, 1, 1, 0, 0, 0, 1, 0, 1)$ that represent the state of the operon being OFF and ON, respectively. 

To test our method we calculated the stationary distribution of the system using Equation~\ref{Eq:StationaryDist} with $g=0.9$ in Equation~\ref{Eq:GoogleMatrix}.
First we used the propensity values given in Equation~\ref{Eq:PropLacten09} where all propensities are fixed to 0.9. This choice of parameters approximates the synchronous dynamics in the sense that each function has a 90\% change of being used during the simulations and 10\% chance of maintaining its current value. 
Under this selection of parameters, the fixed point $s_1$ has a PageRank score of 0.3346 while the other fixed point $s_2$ has a score of 0.0463, see Figure~\ref{lacten09_dist} and Table~\ref{tab:lacten}. Then we have applied our genetic algorithm to search for parameters that can increase the PageRank score of $s_2$ and decrease the score of $s_1$. After doing so, we found the propensity parameters given in Equation~\ref{Eq:PropLacten_prop}. With this new set of parameters, the fixed point $s_1$ has a score of 0.0199 while $s_2$ has a score of 0.5485, see Figure~\ref{lacten_prop_dist} and Table~\ref{tab:lacten}. To appreciate the impact of the change in propensity parameters, we have plotted the state space of the system with both propensity matrices in Figure~\ref{fig:lacten}. The edges in blue in Figure~\ref{fig:lacten} represent the most likely trajectory. Notice that in Figure~\ref{fig:lacten_prop} the trajectories are leading towards $s_2$ and the size of the labels of the nodes were scaled according to their PageRank score, see Table~\ref{tab:lacten}. 
 
\begin{equation}
\label{Eq:PropLacten09}
\begin{tabular}{| c | c  c  c c c  c  c c c c |}
\hline
   & $x_1$ & $x_2$ & $x_3$& $x_4$& $x_5$ & $x_6$ & $x_7$& $x_8$& $x_9$& $x_{10}$\\ \hline
$p_i^\uparrow$& .9 &   .9 &   .9  &  .9 &   .9 &   .9  &  .9 &   .9  &  .9  &  .9\\ 
\hline
$p_i^\downarrow$&.9 &   .9  &  .9 &   .9 &   .9  &  .9 &   .9  & .9  &  .9  &  .9\\ \hline
\end{tabular}
\end{equation}

\begin{equation}
\label{Eq:PropLacten_prop}
\begin{tabular}{| c | c  c  c c c  c  c c c c |}
\hline
   & $x_1$ & $x_2$ & $x_3$& $x_4$& $x_5$ & $x_6$ & $x_7$& $x_8$& $x_9$& $x_{10}$\\ \hline
$p_i^\uparrow$& 0.81 &   1.00 &   0.97  &  0.62 &   0.11 &   0.63  &  0.22 &   0.82  &  0.48  &  0.60\\ 
\hline
$p_i^\downarrow$&0.17 &   0.59  &  0.03 &   0.98 &   0.39  &  1.00 &   0.33  &  0.07  &  0.52  &  0.06\\ \hline
\end{tabular}
\end{equation}

\begin{figure}[h!]
\begin{minipage}[b]{.5\linewidth}
\centering\includegraphics[width=7cm]{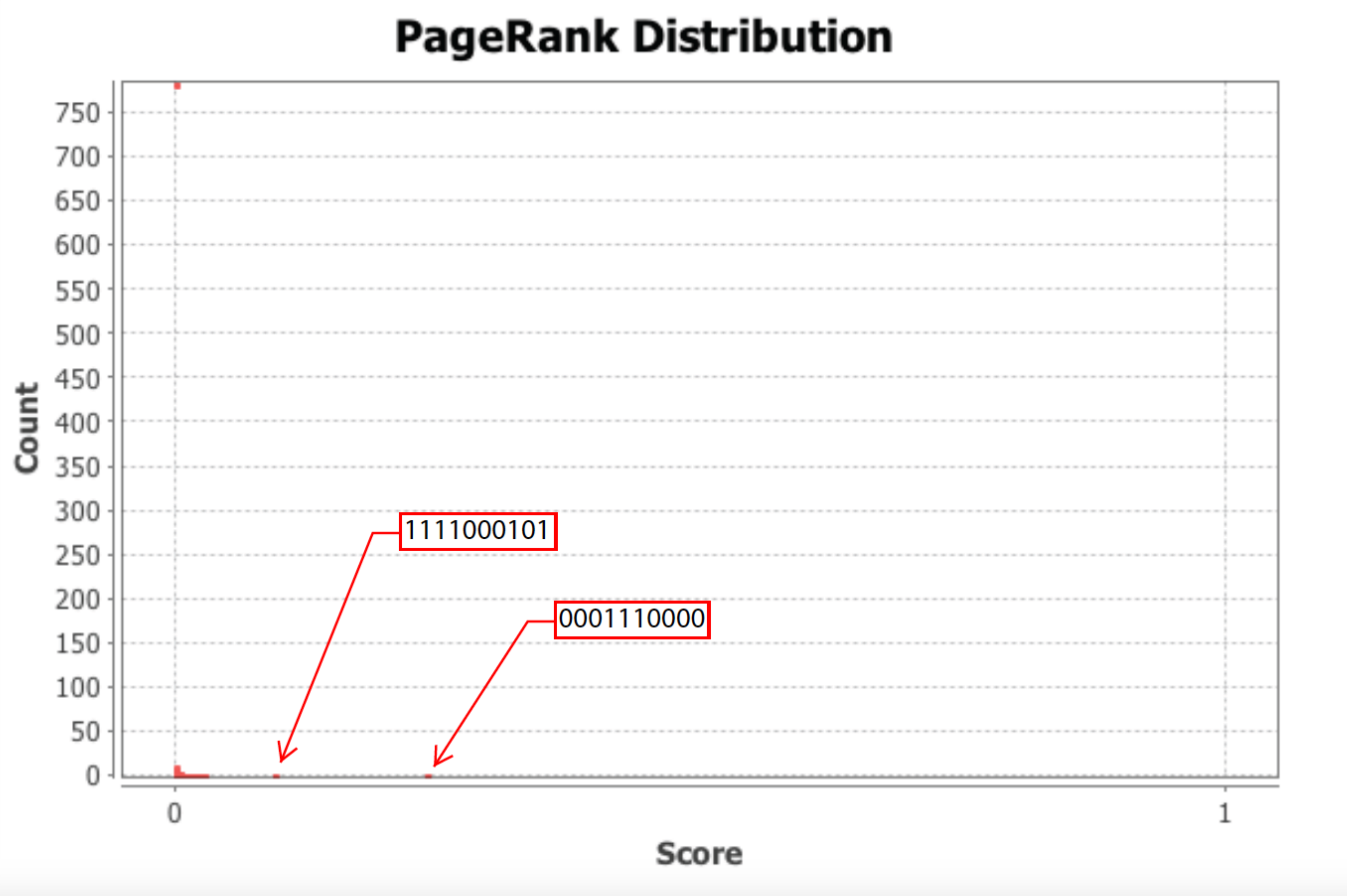}
\subcaption{\tiny{Scores with propensities in Equation~\ref{Eq:PropLacten09}}.}\label{lacten09_dist}
\end{minipage}%
\begin{minipage}[b]{.5\linewidth}
\centering\includegraphics[width=7cm]{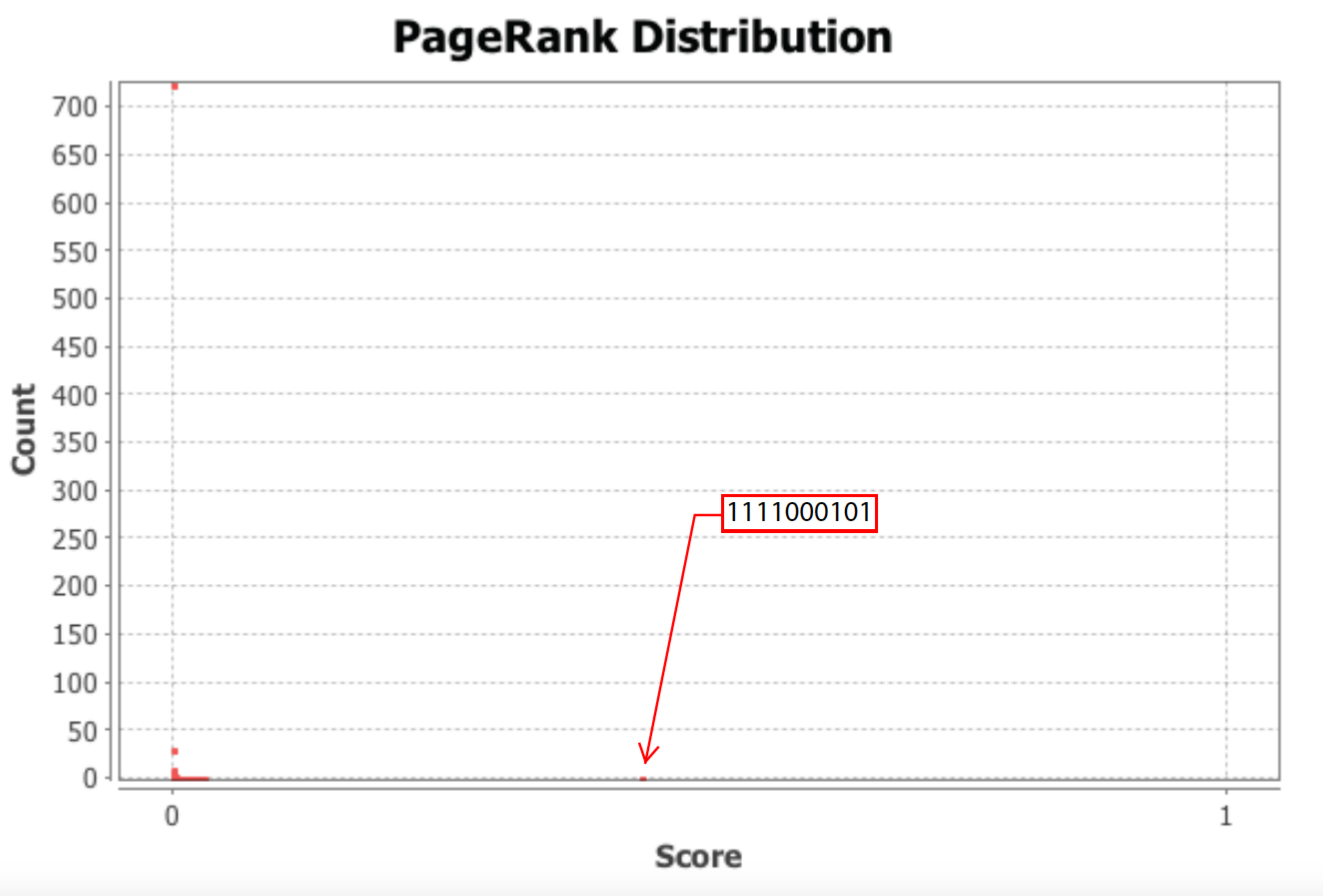}
\subcaption{\tiny{Scores with propensities in Equation~\ref{Eq:PropLacten_prop}}.}\label{lacten_prop_dist}
\end{minipage}
\caption{PageRank scores before and after the genetic algorithm. In each panel, the $x$-axis shows the PageRank scores while the $y$-axis shows the frequencies of states with the given scores in the $x$-axis (the exact scores for the states of interest are given in Table~\ref{tab:lacten}). Left panel shows the state space where all the propensities are equal to 0.9 while the right panel shows the state space where the propensity parameters where estimated using the genetic algorithm.}\label{lacten_dist}
\end{figure}

\begin{figure}[h!]
\begin{minipage}[b]{.5\linewidth}
\centering\includegraphics[width=7cm]{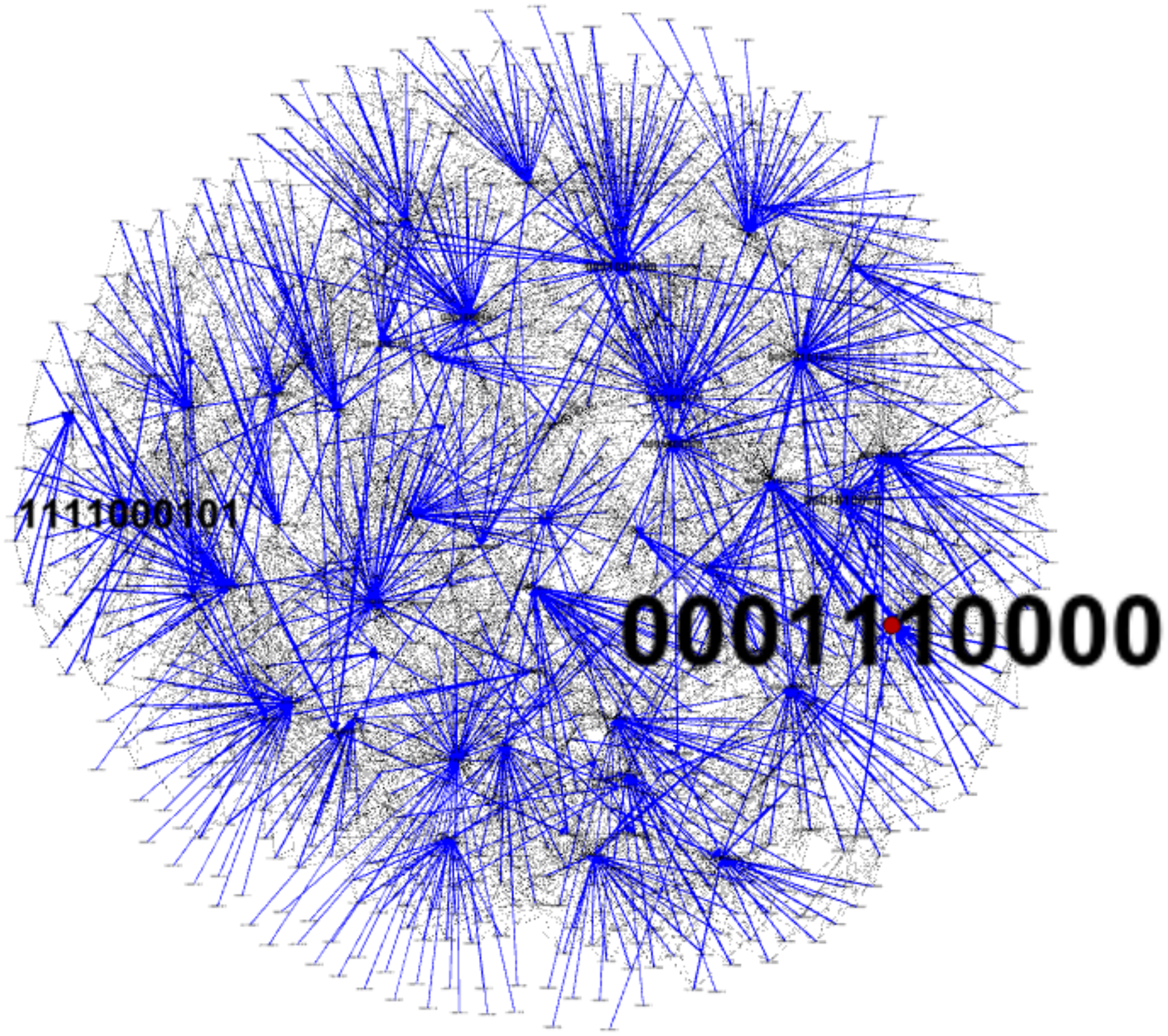}
\subcaption{\tiny{State space with propensities in Equation~\ref{Eq:PropLacten09}}.}\label{fig:lacten09}
\end{minipage}%
\begin{minipage}[b]{.5\linewidth}
\centering\includegraphics[width=7cm]{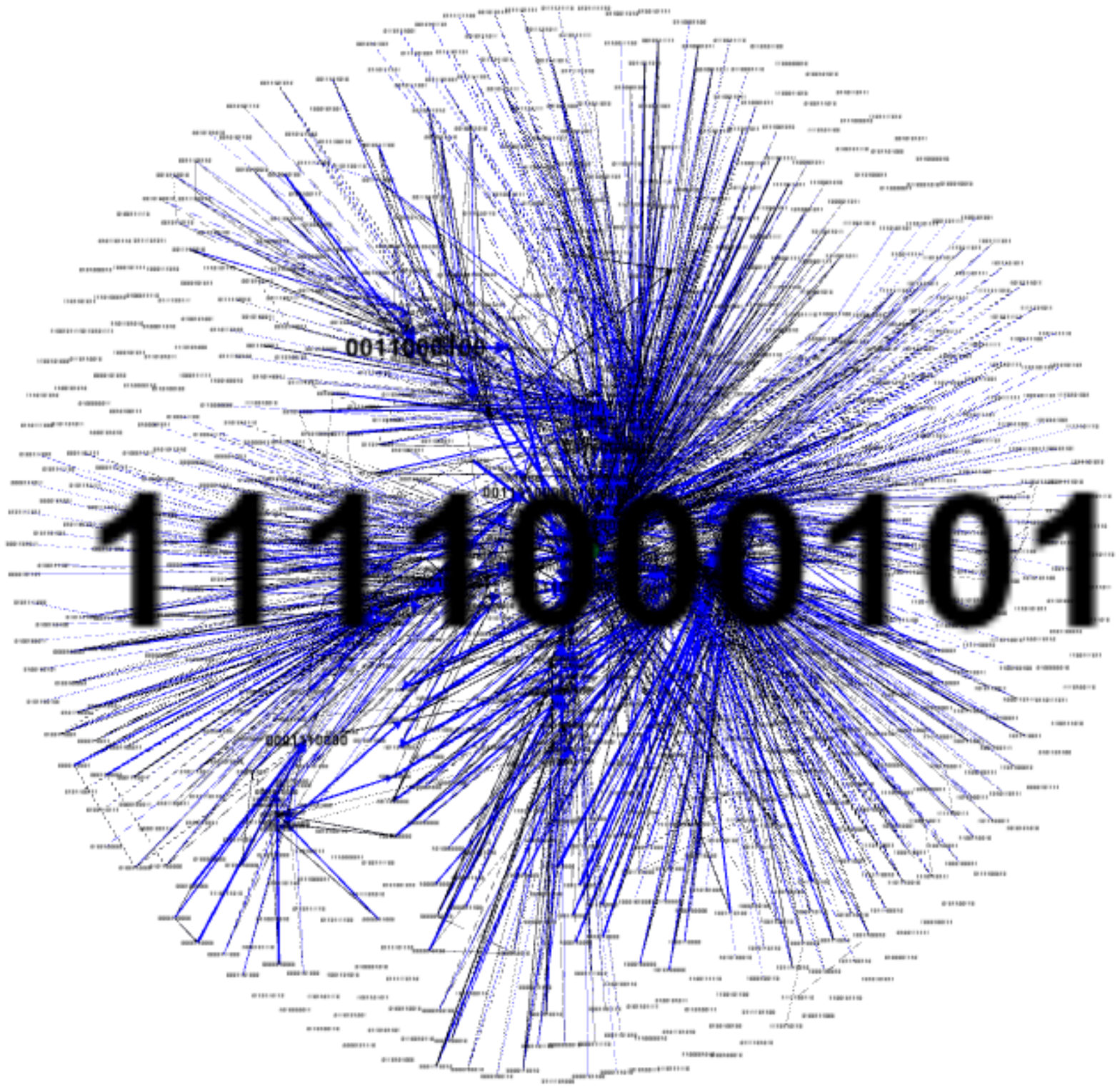}
\subcaption{\tiny{State space with propensities in Equation~\ref{Eq:PropLacten_prop}}.}\label{fig:lacten_prop}
\end{minipage}
\caption{State space comparison before and after the genetic algorithm. Left panel shows the state space where all the propensities are equal to 0.9 while the right panel shows the state space with the estimated propensity parameters using the genetic algorithm. The edges in blue represent the most likely trajectory. The size of the labels of the nodes were scaled according to their PageRank score.}\label{fig:lacten}
\end{figure}

\begin{table}[h!]
  \centering
\begin{tabular}{| l | l | l |}
\hline
 Propensities &Attractor& Score\\ \hline
\multirow{1}{*}{In Equation~\ref{Eq:PropLacten09}}&$0001110000$& 0.3346\\ \cline{2-3}
 (all fixed to 0.9)     &$1111000101$& 0.0463 \\ \hline
   \multirow{1}{*}{In Equation~\ref{Eq:PropLacten_prop}} & $0001110000$& 0.0199 \\ \cline{2-3}
 (genetic algorithm)& $1111000101$& 0.5485\\ \hline 
\end{tabular}
\caption{PageRank scores for the states of the attractors of the system. The order of variables in each vector state is $M,P,B,C,R,R_m,A,A_m,L,L_m$.}
\label{tab:lacten}
\end{table}
\end{example}
\begin{example}\label{Phage lambda infection} \textbf{Phage lambda infection.}
The outcome of phage lambda infection is another system that has been widely studied over the last decades~\cite{ptashne1992phage,Thieffry:1995aa, St-Pierre:2008aa,PMID:20478257,DBLP:journals/ploscb/JohW11}.  
One of the earliest models that has been developed for this system is the logical model by Thieffry and Thomas~\cite{Thieffry:1995aa}. 
 The regulatory genes considered in is this model are: \textit{CI}, \textit{CRO}, \textit{CII}, and \textit{N}. Experimental reports~\cite{Thieffry:1995aa, reichardt1971control, Kourilsky:1973aa, St-Pierre:2008aa} have shown that, if the gene \textit{CI} is fully expressed, all other genes are OFF. In the absence of \textit{CRO} protein, \textit{CI} is fully expressed (even in the absence of \textit{N} and \textit{CII}). \textit{CI} is fully repressed provided that CRO is active and CII is absent. 

This network is a bistable switch between lysis and lysogeny. Lysis is the state where the phage will be replicated, killing the host. Otherwise, the network will transition to a state called lysogeny where the phage will incorporate its DNA into the bacterium and become dormant. These cell fate differences have been attributed to the spontaneous changes in the timing of individual biochemical reaction events~\cite{mcadams1997stochastic, Thieffry:1995aa}. 

In the model of Thieffry and Thomas~\cite{Thieffry:1995aa}, the first variable, $CI$, has three levels $\{0,1,2\}$, the second variable, $CRO$, has four levels $\{0,1,2,3\}$, and the third and fourth variables, $CII$ and $N$, are Boolean. Since the nodes of this model have different number of states, in order to apply our methods, we have extended the model so that all nodes have the same number of states. We have used the method given in~\cite{Veliz-Cuba:2010aa} to extend the number of states such that all nodes have 5 states (the method for extending requires a prime number for number of states so we have chosen 5 states). The method for extending the number of states preserves the original attractors. The update rules for this model are available with our code that is freely available. The extended model has a steady state, $2000$, and a 2-cycle involving $0200$ and $0300$. The steady state $2000$ represents lysogeny where $CI$ is fully expressed while the other genes are OFF. The cycle between $0200$ and $0300$ represents lysis where $CRO$ is active and other genes are repressed.

To test our method we calculated the stationary distribution of the system using Equation~\ref{Eq:StationaryDist} with $g=0.9$ in Equation~\ref{Eq:GoogleMatrix}.
First we used the propensity values given in Equation~\ref{Eq:PropLambda4_09} where all propensities are fixed to 0.9. This choice of parameters approximates the synchronous dynamics in the sense that each function has a 90\% change of being used during the simulations and 10\% chance of maintaining its current value. 
Under this selection of parameters, the fixed point $2000$ has a PageRank score of 0.2772 while the states of the cycle $0200$ and $0300$ have scores of 0.2185 and 0.2108, respectively. Notice that this cycle attractor will have an overall score of 0.4293, see Figure~\ref{fig:PageRankLambda4_09} and Table~\ref{tab:Lambda4}. Then we have applied our genetic algorithm to search for parameters that can increase the PageRank score of the fixed point $2000$ and found the propensity parameters given in Equation~\ref{Eq:PropLambda4_prop}. With this new set of parameters, the fixed point $2000$ has a score of 0.6040 while the states of the cycle $0200$ and $0300$ have scores of 0.0716 and 0.00016, respectively, see Figure~\ref{fig:PageRankLambda4_prop} and Table~\ref{tab:Lambda4}. To appreciate the impact of the change in propensity parameters, we have plotted the state space of the system with both propensity matrices in Figure~\ref{fig:lambda4}. The edges in blue in Figure~\ref{fig:lambda4} represent the most likely trajectory. Notice that in Figure~\ref{fig:lambda4_prop} the trajectories are leading towards $2000$ and the size of the labels of the nodes were scaled according to their PageRank score, see Table~\ref{tab:Lambda4}. 

\begin{equation}
\label{Eq:PropLambda4_09}
\begin{tabular}{| c | c  c  c c |}
\hline
   & $CI$ & $CRO$ & $CII$& $N$\\ \hline
$p_i^\uparrow$& .9 & .9 & .9& .9\\ 
\hline
$p_i^\downarrow$&.9 & .9 & .9& .9\\ \hline
\end{tabular}
\end{equation}
\begin{equation}
\label{Eq:PropLambda4_prop}
\begin{tabular}{| c | c | c | c | c |}
\hline
   & $CI$ & $CRO$ & $CII$ & $N$\\ \hline
$p_i^\uparrow$& 1.0000 & 0 & 0.4277 & 0.7968 \\ \hline
$p_i^\downarrow$& 0.3962 & 1.0000 & 0.6063 & 0.6946 \\
\hline
\end{tabular} 
\end{equation}
\begin{figure}[h!]
\begin{minipage}[b]{.5\linewidth}
\centering\includegraphics[width=7cm]{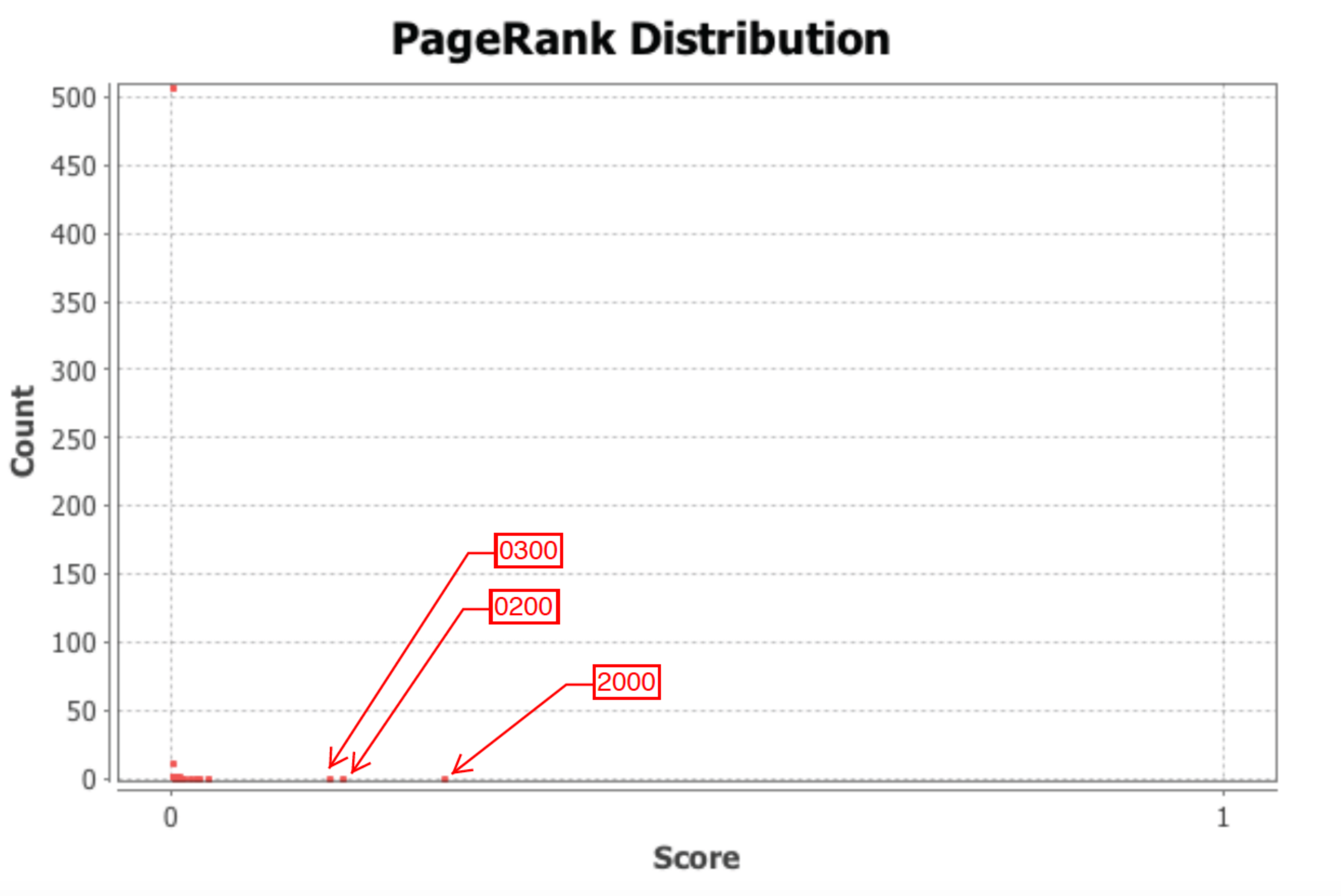}
\subcaption{\tiny{Scores with propensities in Equation~\ref{Eq:PropLambda4_09}}.}\label{fig:PageRankLambda4_09}
\end{minipage}%
\begin{minipage}[b]{.5\linewidth}
\centering\includegraphics[width=7cm]{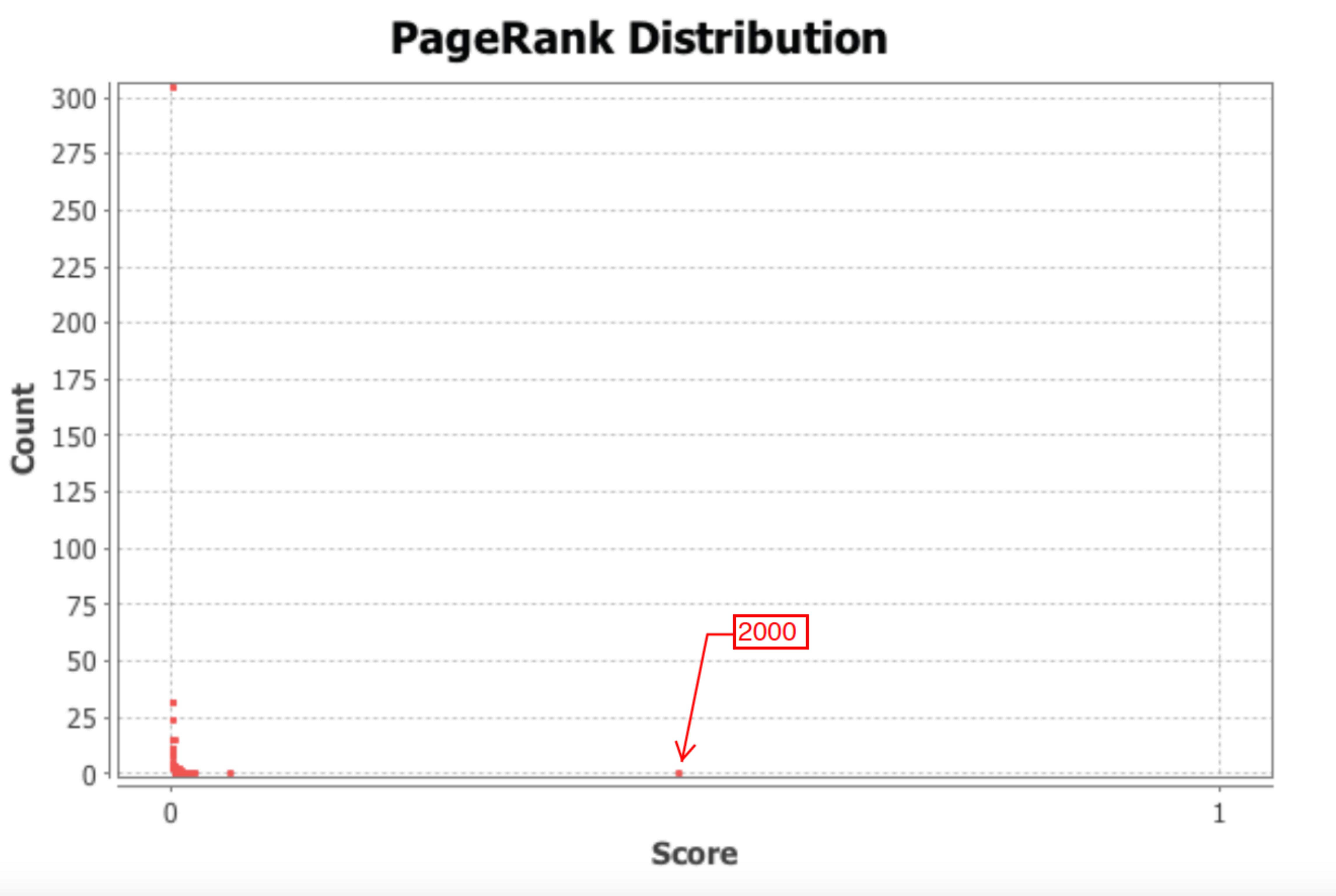}
\subcaption{\tiny{Scores with propensities in Equation~\ref{Eq:PropLambda4_prop}}.}\label{fig:PageRankLambda4_prop}
\end{minipage}
\caption{PageRank scores before and after the genetic algorithm. In each panel, the $x$-axis shows the PageRank scores while the $y$-axis shows the frequencies of states with the given scores in the $x$-axis (the exact scores for the states of interest are given in Table~\ref{tab:Lambda4}). Left panel shows the PageRank scores where all the propensities were equal to 0.9 while the right panel shows the scores where the propensity parameters where estimated using the genetic algorithm. The score for the state $2000$ is 0.6040.}\label{fig:PageRankLambda4}
\end{figure}

\begin{figure}[h!]
\begin{minipage}[b]{.5\linewidth}
\centering\includegraphics[width=6.5cm]{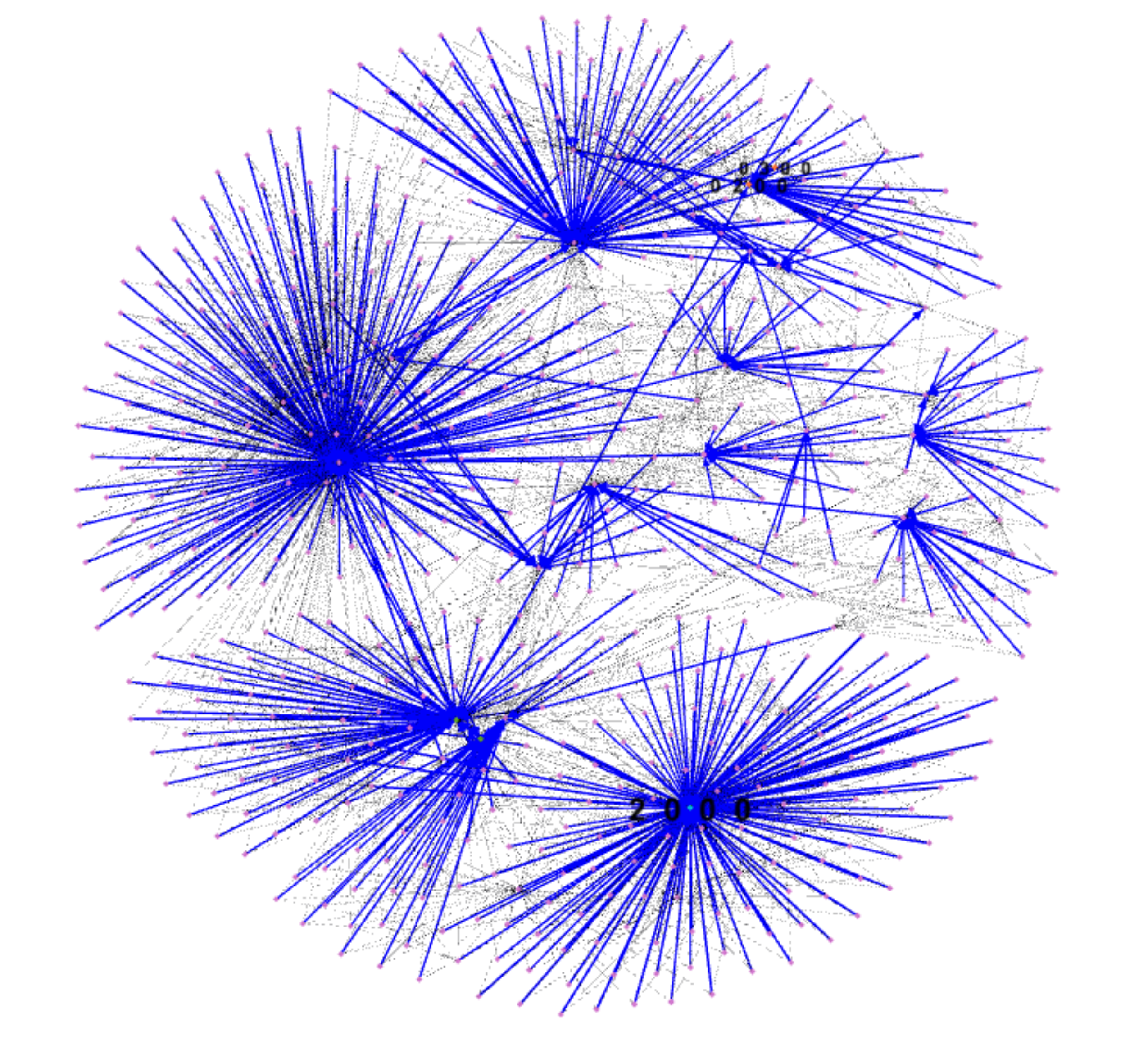}
\subcaption{\tiny{State space with propensities in Equation~\ref{Eq:PropLambda4_09}}.}\label{fig:lambda4_09}
\end{minipage}%
\begin{minipage}[b]{.5\linewidth}
\centering\includegraphics[width=7cm]{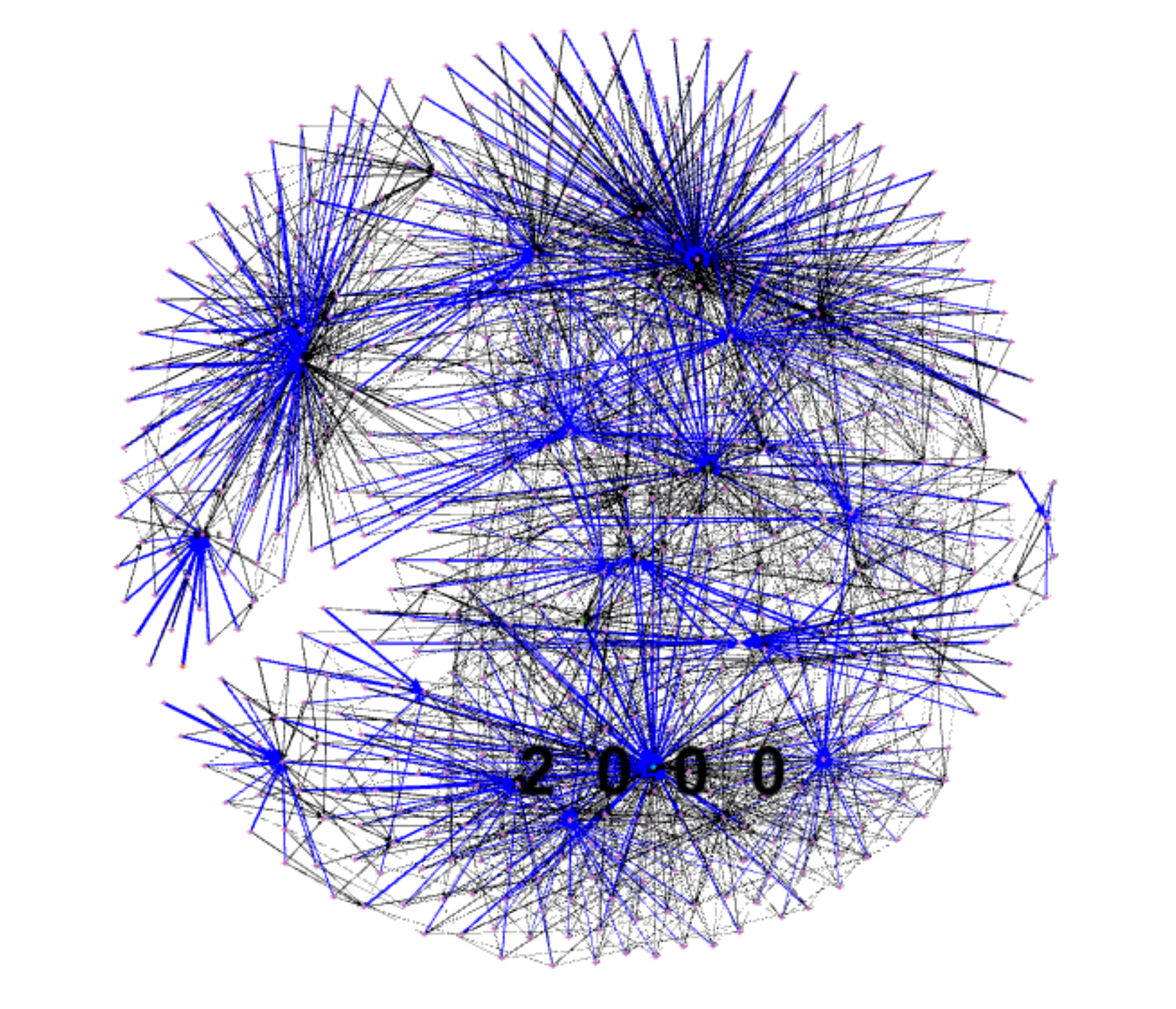}
\subcaption{\tiny{State space with propensities in Equation~\ref{Eq:PropLambda4_prop}}.}\label{fig:lambda4_prop}
\end{minipage}
\caption{State space comparison before and after the genetic algorithm. Left panel shows the state space where all the propensities are equal to 0.9 while the right panel shows the state space with the estimated propensity parameters using the genetic algorithm. The edges in blue represent the most likely trajectory. The size of the labels of the node were scaled according to their PageRank score.}\label{fig:lambda4}
\end{figure}
\begin{table}[h!]
  \centering
\begin{tabular}{| l | l | l |}
\hline
 Propensities &Attractor& Score\\ \hline
\multirow{2}{*}{In Equation~\ref{Eq:PropLambda4_09}}&$2000$& 0.2772\\ \cline{2-3}
                  &$0200$& 0.2185 \\ 
   (all fixed to 0.9)&$0300$& 0.2108 \\ \hline
   \multirow{2}{*}{In Equation~\ref{Eq:PropLambda4_prop}} & $2000$& 0.6040 \\ \cline{2-3}
 & $0200$& 0.0716\\ 
  (genetic algorithm)& $0300$& 0.00016\\ \hline 
\end{tabular}
\caption{PageRank scores for the states of the attractors of the system. The order of variables in each vector state is $CI, CRO, CII, N$.}
\label{tab:Lambda4}
\end{table}
\end{example}
\section{Discussion}
Parameter estimation for stochastic models of biological networks is in general a very hard problem. 
This paper focuses on a class of stochastic discrete models, which is an extension of Boolean networks.
The methods presented here use a well stablished approach for introducing noise into a system in a way that
ergodicity and thus the existence of a unique stationary distribution is guaranteed. Then a genetic algorithm 
for calculating a set of parameters able to approximate a desired stationary distribution was developed.
Also, techniques for approximating the stationary distribution at each iteration of the genetic algorithm that make the search process more efficient was applied.

One shortcoming of the method is that it is a stochastic method. That is, each time we run the algorithm we get a different result. An exhaustive investigation about the variance of the results is still missing. For the examples that we presented in the results section, the output of the algorithm might vary in about 20\% of the reported propensities.

In parameter estimation, sometimes, it is useful to identify a smaller set of key parameters to estimate. This problem is out of the scope of the paper. However, for Boolean network models, one way to address this problem could be by using the different network reduction algorithms, for instance see~\cite{Veliz-Cuba:2014aa,Saadatpour:2010aa}, to identify a smaller ``core" network that preserves the important features of the dynamics of the original network. And then one could apply the parameter estimation techniques that are described in this paper. This type of approach could be especially useful if dealing with very large networks where running the genetic algorithm is computationally expensive.
\section{Conclusions}
In this paper we present an efficient method for estimating the parameters of a stochastic framework. The modeling framework is an extension of Boolean networks that uses propensity parameters for activation and inhibition. Parameter estimation techniques are needed whenever one needs to tune the propensity parameters of the stochastic system to reproduce a desired stationary distribution. For instance, if dealing with a bistable system and if it is desired to have the stationary distribution that have the PageRank score concentrated in one of the attractors of the system, then one needs to estimate the propensity parameters that represent such a desired distribution. Parameter estimation methods for this purpose were not available. In this paper we present a method for estimating propensity parameters given a desired stationary distribution for the system. We tested the method in one Boolean network with 10 nodes (where the size of the state space is $2^{10}=1024$) and a multistate network with 4 nodes where each node has 5 states (where the size of the state space is $4^5=625$). For each system, we were able to redirect the system towards the attractor with the smaller PageRank score. The method is efficient and for the examples we have shown it can be run in few seconds in a laptop computer. Our code is available at \href{http://www.ms.uky.edu/~dmu228/GeneticAlg/Code.html}{http://www.ms.uky.edu/$\sim$dmu228/GeneticAlg/Code.html}.
\section*{Conflict of Interest Statement}
The authors declare that the research was conducted in the absence of any commercial or financial relationships that could be construed as a potential conflict of interest.
\section*{Author Contributions}
DM, JM, and AM designed the project. JM developed the genetic algorithm and implemented the Matlab code. AM run simulations for the results section. DM wrote the paper.
All authors approved the final version of the manuscript.
\section*{Funding}
DM was partially funded by a startup fund from the College of Arts and Sciences at the University of Kentucky.

\section*{Acknowledgments}
The authors thank the reviewers for their insightful comments that have improved the manuscript.


\bibliographystyle{plain}
\bibliography{ref_for_control}      





\end{document}